\documentclass[twocolumn,showpacs,floatfix]{revtex4}

\usepackage{graphicx}
\usepackage{amsmath}
\usepackage{epsfig}
\usepackage{hhline}
\usepackage{amssymb}

\numberwithin{equation}{section}

\newcommand{\av}[1]{\hat{a}_{V,#1}^\dagger}
\newcommand{\bh}[1]{\hat{b}_{H,#1}^\dagger}
\newcommand{\fv}[1]{\hat{f}_{V,#1}^\dagger}
\newcommand{\dv}[1]{\hat{d}_{V,#1}^\dagger}
\newcommand{\fh}[1]{\hat{f}_{H,#1}^\dagger}
\newcommand{\ch}[1]{\hat{c}_{H,#1}^\dagger}
\newcommand{\ev}[1]{\hat{e}_{V,#1}^\dagger}
\newcommand{\ah}[1]{\hat{a}_{H,#1}^\dagger}
\newcommand{\bv}[1]{\hat{b}_{V,#1}^\dagger}
\newcommand{\hv}[1]{\hat{h}_{V,#1}^\dagger}
\newcommand{\cv}[1]{\hat{c}_{V,#1}^\dagger}
\newcommand{\hh}[1]{\hat{h}_{H,#1}^\dagger}
\newcommand{\dhnew}[1]{\hat{d}_{H,#1}^\dagger}
\newcommand{\gv}[1]{\hat{g}_{V,#1}^\dagger}
\newcommand{\gh}[1]{\hat{g}_{H,#1}^\dagger}
\newcommand{\ih}[1]{\hat{i}_{H,#1}^\dagger}

\newcommand{\ph}[1]{\hat{p}_{H,#1}^\dagger}
\newcommand{\pv}[1]{\hat{p}_{V,#1}^\dagger}

\newcommand{\ket}[1]{\left\vert#1\right\rangle}
\newcommand{\beq}{\begin{equation}}
\newcommand{\eeq}{\end{equation}}
\newcommand{\bea}{\begin{eqnarray}}
\newcommand{\eea}{\end{eqnarray}}

\begin{document}

\title{Methods for Producing Decoherence-Free States and Noiseless 
Subsystems Using Photonic Qutrits} 
\author{C. Allen Bishop}
\email{abishop@www.physics.siu.edu}
\author{Mark S. Byrd}
\affiliation{Physics Department, Southern Illinois University, 
Carbondale, Illinois 62901-4401}


\begin{abstract}
We outline a proposal for a method of preparing a single logically encoded 
two-state system (qubit) that is immune to collective noise acting on the 
Hilbert space of the particles supporting it. The logical qubit is 
comprised of three photonic 3-state systems (qutrits) and is generated 
by the process of spontaneous parametric down-conversion. The states 
are constructed using linear optical elements along with three 
down-conversion sources, and are deemed successful by the simultaneous 
detection of six events.  We also show how to select a maximally 
entangled state of two qutrits by similar methods.  For this maximally 
entangled state we describe conditions for the state to be 
decoherence-free which do not correspond to collective errors, but 
which have a precisely defined relationship between them.  
\end{abstract}


\maketitle




\section{Introduction}

Quantum information and quantum technologies have been under active 
investigation for their practical benefits for quite some time.  However, 
only fairly recently have researchers begun to investigate more thoroughly 
the properties and benefits of using higher dimensional Hilbert spaces 
for these purposes.  Higher dimensional systems (in analogy with the term 
qubit for two state systems, d-state systems 
are hereafter referred to as qudits) have properties which are 
quite different from their two-state counterparts which could be useful 
for quantum information processing.  For example, we note that 
$d$-state systems can be more entangled than qubits 
\cite{Caves/Milburn:99,Rungta/etal:00,qutritent}
and can share a larger fraction of their entanglement 
\cite{Wootters:dqudits}.  
We also note that the monogamy rule is different, but 
not well understood for qudits \cite{ou:07}.  
These properties, as well as the larger 
dimension alone, could aid many quantum information processing tasks, 
including quantum key distribution 
\cite{Pasquinucci/etal:June2000,Pasquinucci/etal:Oct2000,
Bourennane/etal:01,Bruss/etal:02}, quantum 
bit commitment \cite{Spekkens/etal:01,Langford/etal:04}, quantum computing 
\cite{Bartlett/etal:02,Klimov/etal:03,Durt/etal:03,Ralph/etal:07}, 
and quantum games.  They also seem to be required, in some form, 
for the solution to a version of the Byzantine agreement 
problem \cite{Fitzi/etal:01}.

As the theoretical benefits become more well-known, proof-of-principle 
experiments will help in our understanding of the behavior of qudits for 
quantum information processing as well as a better understanding of quantum 
mechanics itself.  In this regard, we note several interesting experiments 
using qudits. In the context of quantum computing nuclear 
magnetic resonance techniques have been used to process information 
encoded in single-qutrit systems \cite{Das/etal:03}.  
Qudits have also been used in quantum cryptography to improve reliable 
detection of eavesdroppers \cite{Walborn/etal:06,Groblacher/etal:06}, 
and to demonstrate the 
tossing of quantum coins \cite{Terriza/etal:03}.  

In large part, the differences between qubits and higher 
dimensional systems stems from the difference in the representation theory 
of the group which governs their closed system evolution.  For a 
$d$-state system this group can 
be taken to be SU($d$) if we neglect an overall phase.  
Fundamental differences include entanglement 
properties as well as positivity conditions.  
(It is known that the two are related, see 
\cite{Horodeckis:rmp}.) Such conditions indicate 
important aspects of various entanglements between systems and environments 
\cite{Pechukas+Alicki:95,Jordan:04}.  
This could be vital for ensuring reliable quantum information processing 
such as error correction for open systems which are initially coupled with 
their environment in some way \cite{Shabani/Lidar:06}.  
It is also clear that these 
differences have implications for cryptography, shared reference frames 
\cite{Verst/Cirac:03,Bartlett/Wiseman:03,Mayers:02,Kitaev/etal:ss,Bartlett/etal:04,Bartlett/etal:07}, 
decoherence-free subspaces (DFSs), and noiseless subsystems (NSs) 
\cite{Zanardi:97c,Duan:98,Lidar:PRL98,Knill:99a,Kempe:00,Lidar:00a,Byrd:06} 
(see also \cite{Lidar/Whaley:03,Byrd/etal:pqe04} for reviews), 
through the difference in selection rules for these systems.  

In this paper, we aim to provide a method for producing, albeit indirectly, 
entanglement between three qutrits.  One particular type of entangled state 
we present forms a noiseless state which is protected against noises 
which are not collective and another is a NS protected 
against collective errors.  This subsystem is the smallest subsystem of 
three qudits which can be formed which is immune to collective noise on the 
set of qudit states.  For simplicity, we use qutrits ($d = 3$) 
though in principle any $d$ could be used.  To our knowledge, this is the 
first time an experimental realization of a noiseless subsystem of 
three qudits has be proposed.  Our experiment uses down converted photons and 
linear optical elements to form our NS, the realization of which relies on 
the simultaneous detections of six photons.  This provides a proof-of-principle 
experiment for the formation and manipulation of quantum optical entangled 
qutrits.  

Section \ref{sec:dfstheory} describes the logically encoded states 
which protects information from collective 
errors.   Section \ref{sec:polqutrits} describes the use of the 
polarization of photons to encode qutrit states.  We then describe, 
in Section \ref{sec:stateprep}, methods of selecting these DFS and NS 
states using experimental arrangements consisting of nonlinear crystals, mirrors, 
wave plates, beam splitters and detectors.  
We conclude with a discussion in Section \ref{sec:concl}.


\section{Encoding Logical States}

\label{sec:dfstheory}

In this section we discuss the encoding of logical 
decoherence-free states and noiseless subsystems using 
standard computational basis states.  The corresponding 
physical states are described in the following section.  


\subsection{Encoding into a noiseless subsystem}

In \cite{Byrd:06} it was shown that the Hilbert space of three 
physical qutrits can be used to support a single decoherence-free 
qubit. Each logical state of this decoherence-free qubit is given 
by a superposition of eight 
states labeled $\psi_i^{8,j}$, where the superscript $j$ is a degeneracy 
label used to distinguish the two logical values, i.e., $j = 0,1$. 
The constant superscript $8$ reflects the dimension of 
the irreducible representation of $SU(3)$, and 
the subscript $i$ labels a particular basis state in the 
representation. Using this notation the general form of 
the logical zero state is given by 
\beq
\label{eq:generalZero}
\ket{0_L} = \sum_{i}\alpha_{i}\psi_i^{8,0}.
\eeq
Similarly, the most general logical one state is given by
\beq 
\label{eq:generalOne}
\ket{1_L} = \sum_{i}\beta_{i}\psi_i^{8,1}.
\eeq
The initialization itself is arbitrary as described in Ref.~\cite{Shabani/Lidar}.  
According to the theory of noiseless subsystems, these sets of 
states $\psi_i^{8,0}$  ($\psi_i^{8,1}$) will mix with each 
other, but not with $\psi_i^{8,1}$ ($\psi_i^{8,0}$) under collective 
noise.  As long as this condition holds, the information encoded 
in $\ket{\psi_L} =a \ket{0_L} + b\ket{1_L}$ will be protected.  
The simplest form of logical zero 
results when all but a single expansion coefficient are set to zero. If 
we let $\alpha_{i'}$ remain, logical zero takes the form
\beq
\label{eq:reducedZero}
\ket{0_L} = \alpha_{i'}\psi_{i'}^{8,0}.
\eeq
Suppose that $i' = 3$, then, after normalizing this state 
($\alpha_{3} = 1$), logical zero becomes
\beq
\label{eq:simpleZero}
\ket{0_L} = \psi_{3}^{8,0}.
\eeq
The complete set of 
eight basis states of logical zero $\psi_i^{8,0}$ were given explicitly 
in \cite{Byrd:06} in terms of five quantum numbers. After translating these 
states into a computational basis of the the three-level system, where 
the three orthogonal states are described by the kets $\ket{0}, 
\ket{1}$, and $\ket{2}$, the expression for our example 
can be taken to be
\beq
\label{eq:simpleZeroExplicit}
\ket{0_L} = (\ket{011} - \ket{101})/\sqrt{2}
\eeq                
where we have used, and will continue to use, the shorthand 
notation $\ket{ABC}$ for the 
tensor product of the three qutrit states $\ket{A} \otimes \ket{B} 
\otimes \ket{C}$. In a similar fashion the logical one state 
can be observed to take a simple form by choosing to let all 
but one of the expansion coefficients $\beta_{i}$ vanish. If we 
choose the non-zero coefficient to be $\beta_{8}$, then, after 
normalization, the logical one state 
is given explicitly by
\beq
\label{eq:simpleOneExplicit}
\ket{1_L} = (-\ket{021} + \ket{120} -\ket{201} +\ket{210})/2.
\eeq
For initialization of the logical qubit state, we may create an 
arbitrary superposition of these two logical 
basis states.  After undergoing collective 
decoherence effects, the state $\ket{0_L}$ ($\ket{1_L}$) 
will be of the form Eq.~(\ref{eq:generalZero}) 
(Eq.~(\ref{eq:generalOne})).  


\subsection{Encoding Maximally Entangled States}

\label{sec:singletth}

It was noted and discussed in \cite{Byrd:06} that the two 
inequivalent fundamental representations 
of SU(d), d$\geq$3 
can have a fundamentally different impact on the decoherence-free 
subspaces and noiseless subsystems which they encode.  The 
simplest example exhibiting such a difference is the combination 
of two qutrits of different types.  In the case that one of the 
two inequivalent irreducible representations (irreps) is of 
one type and the other of the other type, a singlet state may be 
formed.  In the case that they are both the same, a singlet 
is not readily available.  This is due to the fact that the 
two different representations are conjugate to each other and 
thus transform in ``opposite'' ways.  

More specifically, in the notation of \cite{Byrd:06}, a singlet 
state of two qutrits can be represented as
\beq
\label{eq:singlet2}
\ket{\Phi_s} = \frac{1}{\sqrt{3}}(\ket{0\bar{0}} + \ket{1\bar{1}} + 
                                    \ket{2\bar{2}}).
\eeq
In this case, the transformation properties are such that one 
of the qutrits will transform according to the conjugate 
representation of the other.  Parameterizations of these two different 
transformations were given as 
$D^{(0,1)}$ and $D^{(1,0)}$ in \cite{Byrd/Sudarshan}.  

Alternatively, the barred states 
could be comprised of two unbarred states.  That is, they may 
arise from an entangled state of two unbarred representations.  
In this case, the state would appear as the totally antisymmetric 
state of three qutrits,
\beq
\label{eq:singlet3}
\ket{\Phi_s} = \frac{1}{\sqrt{6}}(\ket{012} + \ket{120} +  \ket{201} 
                                  - \ket{102} - \ket{021} - \ket{210}),
\eeq
and the state is invariant under collective transformations, 
that is, transformations which act the same on each of the three 
qutrit states.  
(It is interesting to note that this particular state of three 
qutrits has cast doubt on the monogamy relation for qudits \cite{ou:07}.)  

In a given experiment which produces a two qutrit state, the 
transformation properties may not be important.  It could be that we only 
want to produce a particular state in the Hilbert space of two qutrits.  
Here, our objective is to describe experiments which enable the 
selection and exploration of qutrit states.  We therefore describe a 
method of selecting two qutrit states in Section \ref{sec:singlet} 
before discussing the transformation properties.  

Both the logical qubit as well as the singlet state 
have now been encoded in terms of the computational basis states 
of three qutrits. This encoding is, however, quite generic. As 
mentioned above, these three orthogonal states in which a qutrit can
be described are simply referred to as $\ket{0}, \ket{1}$, and $\ket{2}$. 
In the next section we will establish a connection between these generic 
labels and the polarization states of a pair of photons.


\section{Qutrit Encoding Using Photon Polarization}

\label{sec:polqutrits}

The preparation of three-level optical quantum systems has been demonstrated 
in \cite{Bogdanov/etal:04} using the polarization states 
of down-converted photon pairs. There, incident photons with frequency $\nu$  
entering a nonlinear $\beta$-barium borate (BBO) crystal spontaneously 
split into two entangled photons each of which having a 
frequency of roughly $\nu/2$ (conservation of energy 
ensures that the frequencies of the down-converted photons 
sum to the frequency of the pump photon, selection of pairs 
with common frequency can be accomplished by placing narrow-band frequency 
filters outside of the crystal) and linearly polarized either 
along the same axis or along 
orthogonal axes depending on how the crystal is cut. 
Two orthogonal polarization modes can be constructed 
using the former polarization state, corresponding 
to type-I phase matching conditions, by a physical rotation of the crystal 
by $90^{\circ}$. The latter polarization state of the down-converted biphotons 
results from passage through a type-II phase matched crystal. 

Twin photons of equal wavelength produced in a type-I crystal have collinear 
polarization vectors and thus propagate through the crystal 
with equal group velocities 
and emerge from the crystal on a cone due 
to the conservation of transverse momentum. Spatial mode selection of a pair 
may be obtained by placing optical fibers or small apertures 
in a screen on regions of the cone 
that are on opposite sides 
of the pump beam. In type-II down-conversion photon pairs are orthogonally 
polarized and experience different refractive indices 
inside the crystal. Their 
respective group velocities are not equal and they 
emerge on two separate cones, one 
cone for each of the two orthogonal polarizations. By adjusting
the angle between the crystal optic axis and the pump beam the authors in 
Ref. \cite{Kwiat/etal:1995} demonstrated how the 
cones can be made to overlap,
thereby producing entangled Bell-like states along the two directions 
of cone intersection. Along these directions the light 
can be described by the state 
\beq
\label{eq:type2general}
\ket{\Psi} = (\ket{HV} + e^{i\alpha}\ket{VH})/\sqrt{2},
\eeq                 
where $\ket{H}$ and $\ket{V}$ indicate horizontal and vertical polarization, 
respectively. The relative phase $\alpha$, arising from the 
crystal birefringence, 
can be arbitrarily chosen by an appropriate crystal rotation or by 
using an additional 
birefringent phase shifter \cite{Kwiat/etal:1995}. All four Bell 
states were realized as special cases of this general state, in 
particular, the states
\beq
\label{eq:type2special2}
\ket{\psi^+} = (\ket{HV} + \ket{VH})/\sqrt{2},
\eeq
and
\beq
\label{eq:type2special}
\ket{\phi^+} = (\ket{HH} + \ket{VV})/\sqrt{2},
\eeq 
relevant for our purposes here, were 
obtained by setting $\alpha$ equal 
to zero (for both Eq.'s~\ref{eq:type2special} and \ref{eq:type2special2}) 
and placing a half wave 
plate in one photon path angled to rotate horizontal 
polarization to vertical and 
vice versa (for Eq.~\ref{eq:type2special}). An alternative approach to the 
creation of the same entangled states was demonstrated in 
\cite{Kwiat/etal:1999} by replacing the single 
type-II crystal with two adjacent type-I crystals oriented 
at $90^{\circ}$ with respect to each other.

Suppose now, for simplicity, that we choose the set 
($\ket{HH}$, $\ket{VH}$, $\ket{VV}$) as a basis for the qutrit states. 
In order to connect these basis states with the previously 
labeled basis states $\ket{0}$, $\ket{1}$, and $\ket{2}$ let us 
make the following definitions \cite{note1}:
\bea
\label{eq:basisConnection}
\ket{0} &\equiv& \ket{VV}, \nonumber \\
\ket{1} &\equiv& \ket{VH},            \\
\ket{2} &\equiv& \ket{HH}. \nonumber
\eea
These relations can then be used to encode the logical zero and one 
states of the decoherence-free qubit: 
\bea
\label{eq:polarizationEncodingZeroAndOne}
\ket{0_L} &=& (\ket{V_{1}V_{2}V_{3}H_{4}V_{5}H_{6}} - \ket{V_{1}H_{2}V_{3}V_{4}V_{5}H_{6}})/\sqrt{2},
\nonumber \\
\ket{1_L} &=& (-\ket{V_{1}V_{2}H_{3}H_{4}V_{5}H_{6}} + \ket{V_{1}H_{2}H_{3}H_{4}V_{5}V_{6}} 
\nonumber  \\
&& - \ket{H_{1}H_{2}V_{3}V_{4}V_{5}H_{6}} + \ket{H_{1}H_{2}V_{3}H_{4}V_{5}V_{6}})/2.\nonumber \\
\eea
Each of the three qutrits have now been encoded in terms of the 
polarization states of entangled photons produced via 
down-conversion. The numbers attached to the H's and V's serve 
as spatial mode labels and indicate 
a specific photon having a polarization of the type to which 
it is attached, for example, the state $\ket{H_{1}H_{2}}$ corresponds 
to a qutrit state represented by two horizontally polarized photons 
labeled photon 1 and photon 2, etc. 

Now that a physical encoding has been established, the next step is 
to produce the superpositions given by 
Eqs.~(\ref{eq:polarizationEncodingZeroAndOne}).


\section{DFS State Preparation}

\label{sec:stateprep}

Polarizing beam splitters (PBSs), wave-plates, and photodetectors 
have been widely used for the preparation of quantum states 
\cite{Bouwmeester/etal:1999, Lu/etal:07, Zeilinger/etal:1997} and in the 
implementation of quantum 
gates \cite{Ralph/etal:02, O'Brien/etal:03,Hofmann/etal:02,Pittman/etal:01}. Before 
discussing the details of our proposal, we first review the basic idea, given 
in \cite{Zeilinger/etal:1997}, and the extension \cite{Lu/etal:07} on which it rests.
During our discussion of the experiments we will assume that 
each of the nonlinear crystals being used 
simultaneously emits one pair of photons 
and that our detectors are highly efficient.  We 
will also assume that we can eliminate which-way information 
through, for example, the insertion of appropriate narrow-band filters 
as discussed in \cite{Zeilinger/etal:1997,Bouwmeester/etal:1999}.  
These assumptions will simplify our argument.  


\subsection{Selection of a Four-Particle GHZ state}

\begin{figure}[!ht]
\includegraphics[width=.45\textwidth]{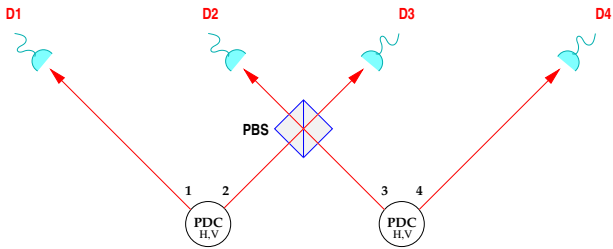}
\caption{(color online). A four-photon polarization-entanglement source \cite{Zeilinger/etal:1997}.}
\label{fig:GHZ}
\end{figure}

Consider the arrangement in Fig.~\ref{fig:GHZ}. 
Ultraviolet light is pumped through two separate 
parametric down-conversion (PDC) sources (e.g. BBO, BiBO crystals) which, upon 
successful down-conversion, emit pairs of photons in the Bell 
state Eq.~(\ref{eq:type2special}), see \cite{note2}. 
The beam-splitter is made to transmit horizontally polarized light 
while reflecting light polarized along the vertical axis. 
Therefore, the photons reaching detector D2 are those from 
path 2 (path 3) having exited the PBS with a
vertical (horizontal) polarization. Similarly, photons reaching detector D3 arrive 
with a vertical (horizontal) polarization after entering the PBS via 
path 3 (path 2).

With a fourfold coincidence detection, one can infer that the 
polarization state of the four-photon system being detected 
is either $\ket{H_1H_2H_3H_4}$ or 
$\ket{V_1V_2V_3V_4}$. The reason is this; in order for detectors D2 
and D3 to fire at the same time one of two photons being detected must have 
come from the left crystal via path 2 and the other from the right crystal 
via path 3, and as mentioned above, if photons coming from paths 
2 and 3 have orthogonal polarizations they 
will both arrive at the same detector thereby ruling out 
a fourfold coincidence 
detection. If detectors D2 and D3 fire simultaneously both 
photons being detected thus get projected into the same polarization 
state which then implies that 
the photons arriving at detectors D1 and D4 must also possess 
that same polarization since 
pairs created in these type-II crystals are emitted in the state 
Eq.~(\ref{eq:type2special}). 
Given our assumptions of path indistinguishability 
\cite{Zeilinger/etal:1997,Bouwmeester/etal:1999}, we will find 
the state $\frac{1}{\sqrt{2}}(\ket{H_1H_2H_3H_4} 
+ \ket{V_1V_2V_3V_4})$.   

The optical arrangement just discussed was extended 
in \cite{Lu/etal:07} to include an additional nonlinear crystal. Using 
three crystals, along with the action of a Hadamard gate, the authors 
just mentioned were able to produce a six-photon ``cluster'' state. 
We will briefly discuss their method in order 
to clarify the details of our proposal.


\subsection{Selection of a Six-Particle GHZ state}

\begin{figure}[!ht]
\includegraphics[width=.45\textwidth]{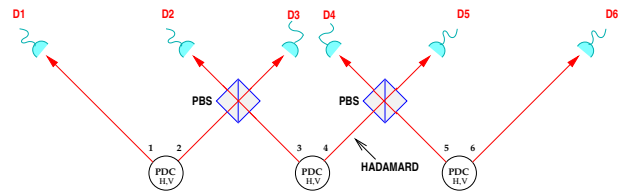}
\caption{(color online). A six-photon polarization-entanglement source.}
\label{fig:GHZ6}
\end{figure}

The setup shown in Fig.~\ref{fig:GHZ6} 
is an extension of the previous 
arrangement which includes an additional nonlinear crystal along with
a Hadamard gate (a wave plate set to rotate the polarization of 
a photon by $45^\circ$) placed in path 4.  
It is not difficult then to convince oneself that in the 
absence of the Hadamard gate, a sixfold coincidence detection projects 
this system into a state that 
can be described by the six-photon GHZ superposition given 
by  $\frac{1}{\sqrt{2}}(\ket{H_1H_2H_3H_4H_5H_6} + \ket{V_1V_2V_3V_4V_5V_6})$. 
In fact, any $n$-photon GHZ state ($n$ = 2, 4, 6, \ldots) of the form 
$\frac{1}{\sqrt{2}}(\ket{H_1H_2 \ldots H_n} 
+ \ket{V_1V_2 \ldots V_n})$ can be selected using the same type of 
extension along with an $n$-fold coincidence detection.
 
\begin{figure}[!ht]
\includegraphics[width=.45\textwidth]{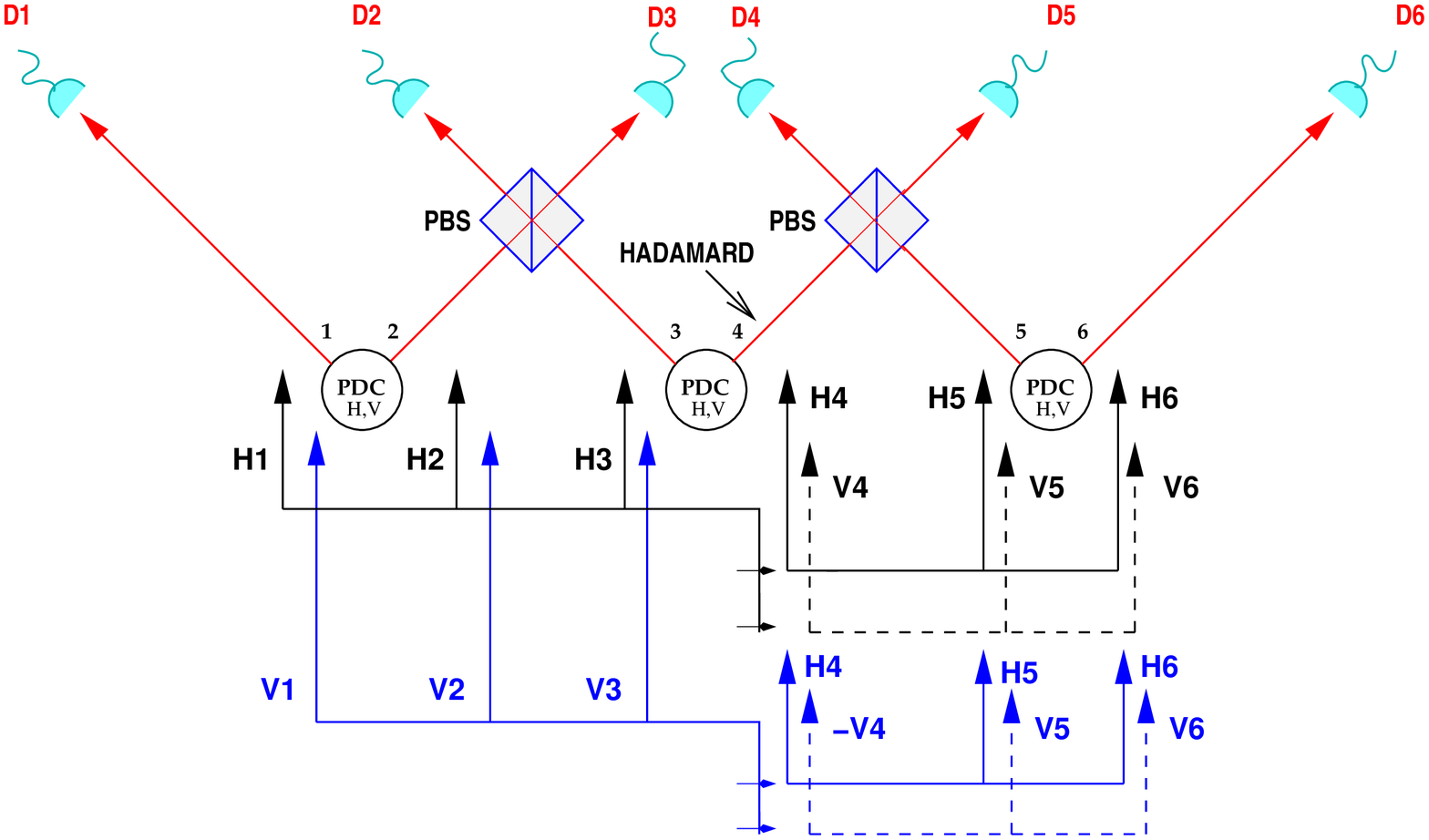}
\caption{(color online). Photon states compatible with six simultaneous detections.}
\label{fig:GHZ6cluster}
\end{figure}

The situation changes when a Hadamard gate $\textbf{H}$ is placed in path 4. 
The action of this gate on the states $\ket{H}$ and $\ket{V}$ is such that 
\bea
\label{eq:HadamardH}
\textbf{H}\ket{H} &=& \frac{1}{\sqrt{2}}(\ket{H} + \ket{V}), \\
\label{eq:HadamardV}
\textbf{H}\ket{V} &=& \frac{1}{\sqrt{2}}(\ket{H} - \ket{V}).
\eea
The inclusion of this gate leads to even more possible 
states that are compatible with a sixfold coincidence 
detection. These new states are shown in 
Fig.~\ref{fig:GHZ6cluster} using the 
lines below the experimental setup. The uppermost line corresponds 
to the left crystal emitting a photon pair which is 
subsequently detected at D1 and D3. In order for D2 to 
fire simultaneously along with the others it must then
detect the presence of a photon emitted into path 3 that exited the 
PBS horizontally polarized. The Hadamard gate in path 4 
rotates the polarization of the photon passing through it so that 
the chances of it being detected at either D4 or D5 are 
equally likely. This Hadamard rotation splits the upper line 
into two equally probable events that are compatible 
with a sixfold coincidence detection. These possibilities can be seen 
by following the two lines, one solid and one dashed, that 
split off from the original.  

The case where the left crystal emits a pair which is 
subsequently detected at D1 and D2 
is represented by the lower set of lines (blue lines) in 
Fig.~\ref{fig:GHZ6cluster}. All four of these possibilities could trigger a 
sixfold detection, and since these photon paths are 
indistinguishable such a detection post-selects 
the six-photon ``cluster'' state
\bea
\label{eq:cluster}
\ket{C_6} &=& \frac{1}{2}(\ket{H_1H_2H_3H_4H_5H_6} 
+ \ket{H_1H_2H_3V_4V_5V_6} \nonumber \\
&& + \ket{V_1V_2V_3H_4H_5H_6} - \ket{V_1V_2V_3V_4V_5V_6}) \nonumber \\
\eea
realized in \cite{Lu/etal:07}. With these 
examples in mind we proceed with 
the preparation of a logical decoherence-free 
qubit encoded in terms of the six-photon GHZ states given in 
Eq. (\ref{eq:polarizationEncodingZeroAndOne}).


\subsection{Physical Preparation of Logical Zero}

\begin{figure}[!ht]
\includegraphics[width=.48\textwidth]{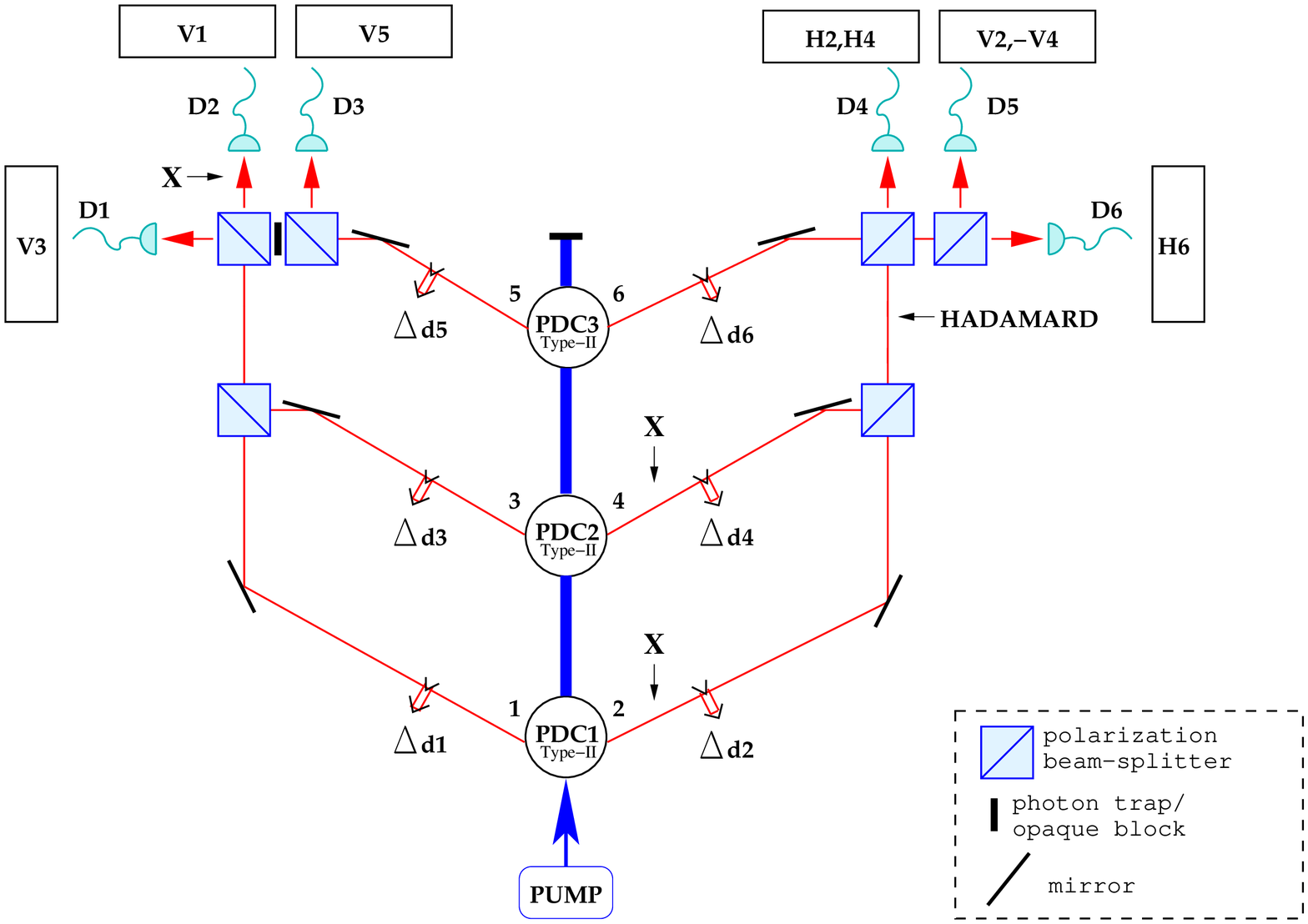}
\caption{(color online). A schematic of a setup that can be used for the 
preparation of a decoherence-free logical zero state. Boxes adjacent to 
detectors D1, D2, $\ldots$ , D6 indicate possible photon states being 
detected in the event of six simultaneous detections. 
The X gates ($X \equiv \sigma_x$ is a wave plate set to rotate 
the polarization by 90 degrees) 
transform $\ket{H}$ into $\ket{V}$ and vice 
versa. The Hadamard gate transforms states according to 
Eq's. (\ref{eq:HadamardH} and \ref{eq:HadamardV}).}
\label{fig:logical0}
\end{figure}

In Fig.~\ref{fig:logical0} we present a pictorial illustration of a setup 
which can be used to prepare the decoherence-free logical zero state given 
in Eq.~(\ref{eq:polarizationEncodingZeroAndOne}). The arrangement 
is similar to that used in \cite{Lu/etal:07} in that 
its photon supply depends on three down-conversion 
sources and in the use of polarization 
beam-splitters to direct orthogonally polarized photons down 
separate paths. Also, successful state preparation again rests on the 
simultaneous detection of six photons, a pair from each of the 
three different crystals. 
The three crystals, PDC's 1, 2, and 3, are each fed 
ultraviolet light by the same source and emit photons in the Bell state 
given by Eq.~(\ref{eq:type2special2}). For convenience, let us refer to the 
photon emitted into the particular path $i$ 
as photon $i$, so, for example, the photon emitted by PDC2 into 
path 3 will be called photon 3, etc. 
Now, occasionally these independent down-conversion processes occur in 
the three crystals at the 
appropriate times needed for six simultaneous detections. If all 
optical path lengths are equal, a sixfold coincidence 
detection requires nearly simultaneous down-conversion 
to take place in each of the three crystals.  
Note that we may assume that all optical paths are equal 
since if they were not, the path lengths can be adjusted 
by by placing mirrors in some or all of the paths and 
varying $\Delta$d1, $\Delta$d2, $\ldots$ , $\Delta$d6.  

Assuming these adjustments have been made 
correctly, and that no down-converted photons have been lost, there 
will be nearly simultaneous recordings from D1, D2, D3, and D6 due to 
the respective detections of photons 3, 1, 5, and 6. This results 
because there are no other paths for them to take (note 
the opaque block serving as a photon trap placed 
between the PBSs adjacent to D2 and D3). This combined 
four photon system is thus detected in the polarization state
\beq
\label{eq:photons1,3,5,6}
\ket{\Psi_{1,3,5,6}} = \ket{V_1V_3V_5H_6}.
\eeq 
On the other hand, if photons 2 and 4 
arrive at either D4 or D5 and both of these detectors fire 
simultaneously it is impossible to know, based 
only on the knowledge of these two detection events, which one of them 
arrived in the horizontal state and which one 
arrived in the vertical state. This limited knowledge implies that the 
state of photons 2 and 4 is 
\beq
\label{eq:photons2,4}
\ket{\Psi_{2,4}} = \frac{1}{\sqrt{2}}(\ket{V_2H_4} - \ket{H_2V_4}).
\eeq 
In the event of six simultaneous detections one can infer that 
the decoherence-free logical zero state 
$\ket{0_L} = (\ket{V_{1}V_{2}V_{3}H_{4}V_{5}H_{6}} - 
\ket{V_{1}H_{2}V_{3}V_{4}V_{5}H_{6}})/\sqrt{2}$ of 
Eq. (\ref{eq:polarizationEncodingZeroAndOne}) has 
been successfully prepared (see Appendix A for more details). 
With a few small modifications, the 
setup just described can also be used for the preparation of the 
logical one state in Eq. (\ref{eq:polarizationEncodingZeroAndOne}). We 
present that arrangement next and discuss the events necessary for 
a sixfold coincidence detection.


\subsection{Physical Preparation of Logical One}

By moving the Hadamard, removing one of the X gates, 
and replacing two of the 
polarized beam-splitters with unpolarized (ordinary) beam-splitters 
the setup shown in Fig.~\ref{fig:logical0} can be modified 
to prepare a decoherence-free logical one state. This arrangement is 
shown below in Fig.~\ref{fig:logical1}.  On the 
left side of this figure we see that if detectors 
D1, D2, and D3 all fire at once the combined photon 1, 3, and 5 
system was detected in the state
\beq
\label{eq:photons1and3and5}
\ket{\Psi_{1,3,5}} = \frac{1}{\sqrt{2}}(\ket{V_1H_3V_5} + \ket{H_1V_3V_5}).
\eeq 
\begin{figure}[!ht]
\includegraphics[width=.48\textwidth]{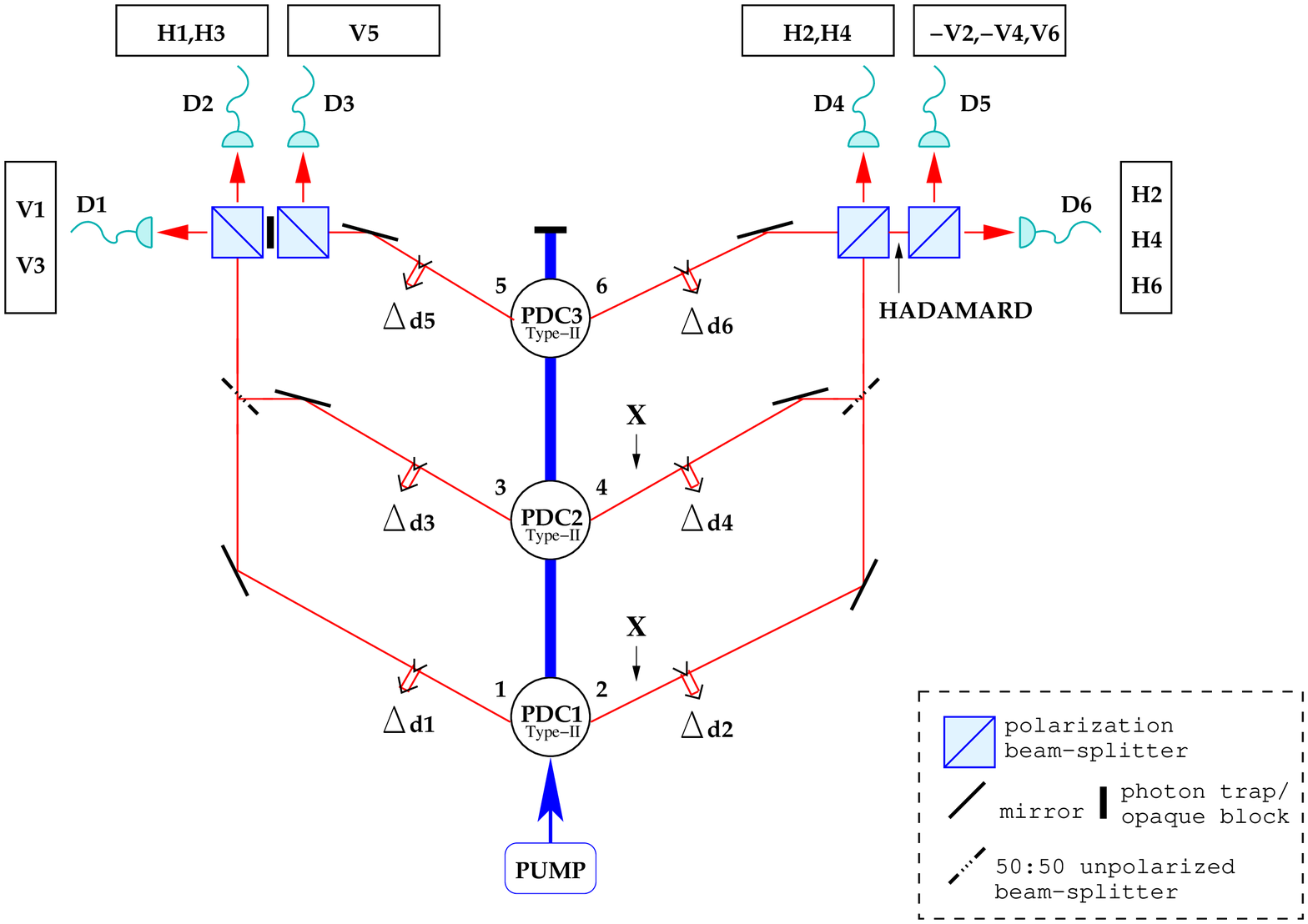}
\caption{(color online). A schematic of a setup that can be used for the 
preparation of a decoherence-free logical one state.}
\label{fig:logical1}
\end{figure}
Consider the case where photons 1, 3, and 5 
are detected in the state $\ket{V_1H_3V_5}$ 
corresponding to the first term in Eq.~(\ref{eq:photons1and3and5}). 
In this case one may infer that photon 2 was 
emitted in a horizontally polarized state 
and photon 4 in a vertical state since there are no 
polarization rotating devices in which photons 1 and 3 encounter. This 
being the case, after a common rotation by separate X gates, 
photons 2 and 4 exit the first PBS via different output 
paths. Now 
photon 4, being horizontally polarized, exits through the top path 
where it is detected at D4. Photon 2, being vertically polarized, exits the 
PBS via the right side and then passes through a Hadamard gate rotating its 
state in accordance with Eq.~(\ref{eq:HadamardV}). After passing through the 
next PBS photon 2 continues on its way 
propagating toward detectors D5 and 
D6. If it happens to trigger D5 photon 6 will then be detected 
at D6 in a horizontal state in the event of 
six simultaneous detections. The six-photon state corresponding to 
this set of events is given by 
\beq
\label{eq:case1}
\ket{\Psi_{1,2,3,4,5,6}^{(1)}} = 
- \ket{V_1V_2H_3H_4V_5H_6},
\eeq
which is indeed one of the terms forming logical one. If photon 2 is 
instead detected at D6 in a horizontal state, a sixfold coincidence detection 
requires photon 6 to be detected at D5 in a vertical state yielding 
the second possibility
\beq
\label{eq:case2}
\ket{\Psi_{1,2,3,4,5,6}^{(2)}} = 
\ket{V_1H_2H_3H_4V_5V_6}.
\eeq
Again, this state can be seen as one of the terms 
forming logical one. Note that these are the only two states that can trigger 
six simultaneous detections when photons 1, 3, and 5 are described by 
$\ket{V_1H_3V_5}$.  However, if photons 1, 3, and 5 are in the state 
forming the second term in Eq.(\ref{eq:photons1and3and5}), 
namely $\ket{H_1V_3V_5}$, two more possibilities arise. The same type of 
argument shows that these states are given by  
\beq
\label{eq:case3}
\ket{\Psi_{1,2,3,4,5,6}^{(3)}} = 
- \ket{H_1H_2V_3V_4V_5H_6},
\eeq
and
\beq
\label{eq:case4}
\ket{\Psi_{1,2,3,4,5,6}^{(4)}} = 
\ket{H_1H_2V_3H_4V_5V_6}.
\eeq
These are the only four states that are compatible with a sixfold 
coincidence detection, the 
setup just described can therefore be used to prepare the 
four-term coherent superposition representing the 
decoherence-free logical one state.


\subsection{Preparation of a Maximally Entangled Two-Qutrit State}

\label{sec:singlet}

As discussed in \cite{Byrd:06}, a maximally entangled state of 
two qutrits can take one of two forms, one given by 
Eq.~(\ref{eq:singlet2}), and one by
\beq
\label{eq:maxent}
\ket{\Phi_m} = \frac{1}{\sqrt{3}}(\ket{00} + \ket{11} + 
                                    \ket{22}).
\eeq
The difference in the barred and unbarred states, as a 
practical matter, is the way in which it transforms.  
The state given in Eq.~\ref{eq:singlet2} 
is decoherence-free while $\ket{\Phi_m}$ is not.  
This is due to the difference in transformation properties.  
The transformation 
matrices will not be repeated here, but were given in 
\cite{Byrd/Sudarshan} in a particular parameterization.

\begin{figure}[!ht]
\includegraphics[width=.50\textwidth]{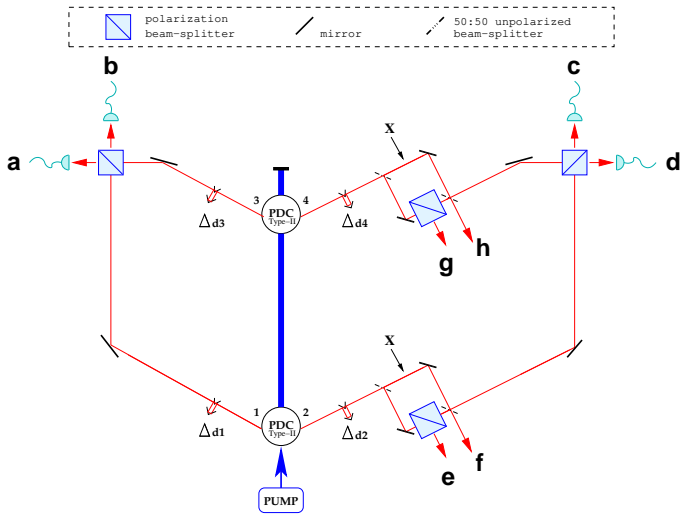}
\caption{(color online). An optical arrangement that may be used for the 
preparation of a maximally entangled two-qutrit state.}
\label{fig:singlet}
\end{figure}

\begin{figure}[!ht]
\includegraphics[width=.25\textwidth]{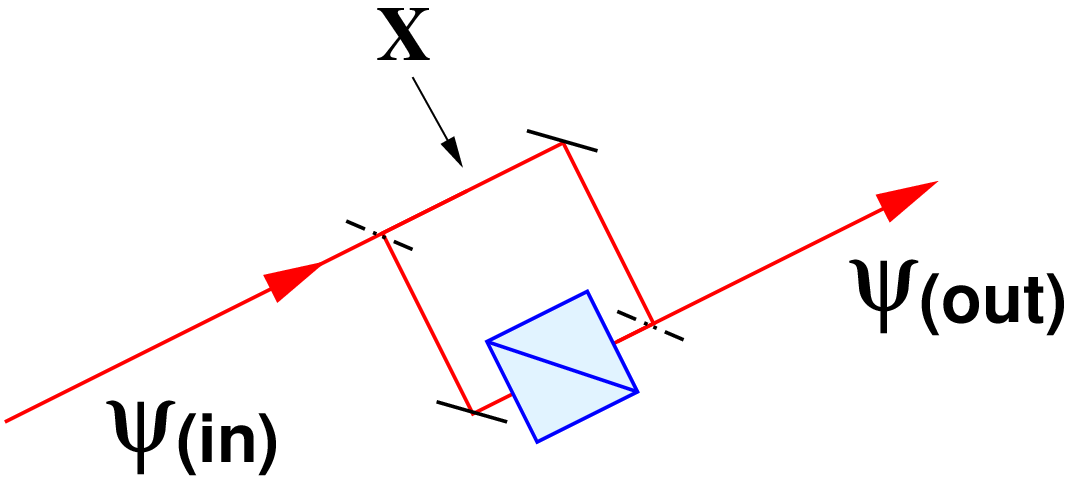}
\caption{(color online). Gate used in the preparation of a maximally entangled 
2-qutrit state. For 
$\ket{\psi(in)} = \ket{H}$ the gate outputs the 
state $\ket{\psi(out)} = 
\frac{1}{2}(\ket{H} + \ket{V} + \sqrt{2}\ket{vacuum})$. For 
$\ket{\psi(in)} = \ket{V}$ the corresponding output is $\ket{\psi(out)} = 
\frac{1}{2}(\ket{H} +\sqrt{3}\ket{vacuum})$.}
\label{fig:singletGate}
\end{figure}

In terms of the definitions above for $\ket{0}$, $\ket{1},$ and 
$\ket{2}$, a maximally entangled state takes the form
\beq
\label{eq:singletPolarization}
\ket{\psi_{max}} = \frac{1}{\sqrt{3}}(\ket{V_{1}V_{2}V_{3}V_{4}} 
+ \ket{V_{1}H_{2}V_{3}H_{4}} + \ket{H_{1}H_{2}H_{3}H_{4}}).
\eeq
This particular state 
may be post-selected based on a fourfold coincidence detection 
using the optical arrangement shown in Fig.~\ref{fig:singlet}. 
Photons emitted into paths 2 and 4 are both probabilistically 
transmitted through identical gates, 
shown in Fig.~\ref{fig:singletGate}, before being detected by 
either c or d.
Since photon pairs are emitted in accordance with 
Eq.~(\ref{eq:type2special2}) the only three possible states 
that can produce four simultaneous detections at a, b, c, and d are 
those forming the superposition of Eq.~(\ref{eq:singletPolarization}). 
The transmission coefficients of the gates (Fig.~\ref{fig:singletGate}) 
on the right for states $\ket{H}$ and $\ket{V}$ ensure that the 
probabilities of each term forming this superposition are equal 
(see Appendix A for more details).

For this state to be decoherence-free, we would require, as 
mentioned in Sec.~\ref{sec:singletth}, each qutrit to 
transform differently, and in some sense they must transform 
in an opposite fashion.  Clearly if they transform in the 
same way, the state $\ket{\Phi_m}$ in Eq.~(\ref{eq:maxent}) 
is changed.  Take for example the transformation 
$$
\ket{j}\rightarrow \exp(-i\alpha_j)\ket{j},
$$
where $\alpha_j$ is different for each $j (=0,1,2)$, for example 
$\alpha_j =\pi j/10$.  In this case the new state is different from 
the original even after an overall phase is extracted and even though 
each have been transformed in the same way.  This is not a 
collective decoherence-free state.  On the other hand, consider 
the simultaneous transformation
$$
\ket{j}\rightarrow \exp(-i\alpha_j)\ket{j},
$$
for the basis states of the first and 
$$
\ket{j}\rightarrow \exp(+i\alpha_j)\ket{j},
$$
for the basis states of the second qutrit.  
In this case $\ket{\Phi_m}$ is unchanged from the original.  
In general, the first qutrit must transform according 
to the conjugate transformation of the second qutrit.  This 
is the sense in which the representations need to be ``opposite'' 
of each other.  The most general such transformations are 
given by $D^{(0,1)}$ and $D^{(1,0)}$ in \cite{Byrd/Sudarshan}.  

Many decoherence-free, or 
noiseless subsystems which have been described theoretically, 
and which have been produced in the laboratory, have been those 
which are immune to collective errors.  
This can be explained, in part, by the simple structure of 
such states as explained in \cite{Byrd:06}.  However, 
here we have 
provided a particular example of a state which is 
decoherence-free under the simultaneous, but {\it not} collective 
transformation of two states.


\section{Conclusion}

\label{sec:concl}

We have presented a proposal for the physical 
preparation of a single decoherence-free qubit embedded in the product 
space of three physical qutrits. Two similar setups 
have been described to complete this task; one for the preparation 
of the logical zero state, and the other for the preparation 
of the logical one state. They are similar up to the inclusion (or 
exclusion) of an X gate, the location 
of a Hadamard gate, and the replacement of two polarized (unpolarized) 
beam-splitters with two unpolarized (polarized) beam-splitters. 
Distinguishing between logical zero and logical one after being 
prepared in this way could prove to be a formidable task 
since the basic constituents of these post-selected states arriving 
at a particular detector (assumed here to be a 
quantum non-demolition device) all 
carry the same polarization and are by necessity 
indistinguishable. Decoding these states and implementing the derived 
logical gates compatible with a three-qutrit NS undergoing 
collective noise \cite{Byrd/Bishop:06, Bishop/Byrd:07} 
are subjects for future research.

We have also presented an experimental arrangement for the 
production of a maximally entangled state of two qutrits, which when 
transformed properly, exhibits decoherence-free effects.  This will 
enable the determination of the experimental difference 
between inequivalent representations of qutrit states.  
Furthermore this experiment provides an example of a 
state which would remain unchanged under the simultaneous, 
but not identical, transformations of its constituents.  

A possible downside to our proposed experiments is that they 
produce the desired states only in the case of fourfold 
(for the singlet) and sixfold (for the logical states) 
coincidence detections, which might be quite 
infrequent. The authors in \cite{Lu/etal:07} have realized such 
sixfold coincidence detections but the rate at which they occurred 
remains unclear. If this hurdle is not too high the 
proposed proof-of-principal experiment could be extended for practical use by 
using quantum non-demolition (QND) measurements 
at each of the six outputs. In \cite{Kok/etal:02} 
a proposal for such QND devices was given that uses only beam-splitters 
and photo-detectors along with auxiliary photons.  
Another concern is due 
to the fact that a $n$-pair emission in a single crystal is nearly 
as probable as $n$ down-conversions taking place in separate 
crystals \cite{Walther/etal:04,Pan/etal:01,Simon/Pan:02}. 
If we restrict our attention for the moment to the 
case of only three simultaneous down-conversion processes occurring, 
we see that the arrangement used to prepare the logical zero state 
only yields sixfold detections in the event 
of single-pair emissions in each of the 
crystals. On the other hand, the arrangement used to 
prepare the logical one state can produce a sixfold detection when 
a double-pair is produced in one of the crystals while no pairs 
are produced in another. This apparent problem vanishes if we recognize 
the fact that the numbers attached to specific photons serve merely as 
labels indicating which particular spatial mode a particular photon 
was emitted into. If two down-conversions happen to take place in 
PDC1, for example, while no photons are converted in PDC2, the arrangement 
can still post-select the appropriate superposition based on six 
detections as long as we label photons properly. 
This same reasoning applies to the arrangement used to 
prepare the maximally entangled 
state in the case of double emissions in one of 
the two crystals and none in the other. If we extend our analysis 
to situations where more than three down-conversions take place 
simultaneously we cannot address the problem by simple label considerations 
since then there might be more than six photons arriving at the detectors. 
If the detectors were number resolving this would not be an issue 
since we could post-select states based on the arrival and detection 
of six and only six photons at the appropriate detectors. Although 
the probability of witnessing higher numbers of photon pairs created 
simultaneously dramatically decreases with increasing pairs, 
these problems could be eliminated in the future if improvements in 
entangled photon sources like 
those presented in \cite{Stevenson/etal:06} are made.

Path-indistinguishability is also essential for the 
preparation of quantum superpositions. When 
a single pump photon decays into a pair of daughters they emerge 
highly correlated in frequency due to the conservation of energy, and their 
``spontaneous'' emission clearly signifies a highly temporal correlation
as well. These correlations, along with others, could in principal 
be exploited 
to determine which photon pair arrived at which detectors. In order to 
account for these possible exploitations the authors 
in Ref.~\cite{Zeilinger/etal:1997} suggested placing narrow-band 
filters, centered at half of the pump frequency, in the photon paths. 
This would not only minimize the frequency 
differences of detected photons, but also reduce the temporal correlations 
since there is an uncertainty in the 
time in which it takes to pass through one of them. 
If the uncertainty in the amount of time in which 
it takes to pass through a filter was high enough to 
establish an intrinsic uncertainty in the 
arrival times of photons belonging to a pair yet still remained compatible 
with the resolution time of a detector so that six nearly simultaneous 
detections could occur, the condition of path-indistinguishability could be 
realistically met.  

We believe these proposed experiments are able to be performed using 
readily available technologies and will allow the exploration of 
new types of qutrit states--decoherence-free subspaces and noiseless 
subsystems comprised of qutrit states.

NOTE: After the completion of this work, several authors have presented 
related experiments which help show the feasibility of our 
proposed experiments \cite{Lanyon/etal:07,Vallone/etal:07}.  


\appendix


\section{Detailed Calculations of Mode Transformations}

\begin{figure}[!ht]
\includegraphics[width=.50\textwidth]{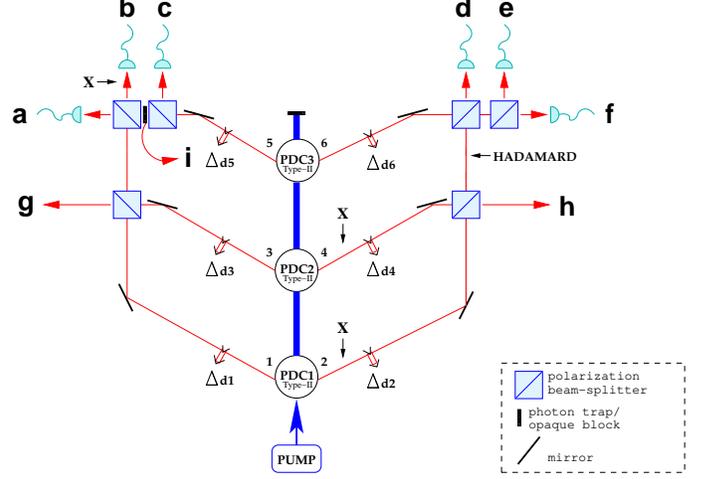}
\caption{(color online). An optical arrangement that may be used for the 
preparation of a decoherence-free logical zero state.}
\label{fig:logical0modes}
\end{figure}

The states post-selected by coincidence detections at the 
appropriate outputs of the optical arrangements shown in this paper 
may be calculated in terms of the transformations of the 
down-converted modes. If we let $\ph{i}$, and $\pv{i}$ 
respectively represent the creation operators 
for the horizontal and vertical polarization modes of the $i^{th}$ 
photon ($i = 1, 2, 3, 4, 5, 6$) then a successful down-conversion in PDC1 
may be described by the operator $\hat{O}_{1,2}^\dagger$, 
where $\hat{O}_{1,2}^\dagger$ is given by
\beq
\label{eq:creationOperatorForBB01}
\hat{O}_{1,2}^\dagger = (\ph{1}\pv{2} + \pv{1}\ph{2})/\sqrt{2}.
\eeq
Similarly, a successful down-conversion in PDC2 (PDC3) can be described by 
the creation operators $\hat{O}_{3,4}^\dagger$ 
($\hat{O}_{5,6}^\dagger$), where 
$\hat{O}_{3,4}^\dagger$ ($\hat{O}_{5,6}^\dagger$) are given by
\beq
\label{eq:creationOperatorForBB02}
\hat{O}_{3,4}^\dagger = (\ph{3}\pv{4} + \pv{3}\ph{4})/\sqrt{2},
\eeq
and
\beq
\label{eq:creationOperatorForBB03}
\hat{O}_{5,6}^\dagger = (\ph{5}\pv{6} + \pv{5}\ph{6})/\sqrt{2}.
\eeq
The creation operator corresponding to the event of 
two independent down-conversion processes occurring 
simultaneously (one in PDC1 and another in PDC2) is thus given by 
\bea
\label{eq:twoDownConversions}
\hat{O}_{1,2}^\dagger\hat{O}_{3,4}^\dagger &=& 
(\ph{1}\pv{2} + \pv{1}\ph{2})(\ph{3}\pv{4} + \pv{3}\ph{4})/2 \nonumber \\
&=& (\ph{1}\pv{2}\ph{3}\pv{4} + \ph{1}\pv{2}\pv{3}\ph{4} \nonumber \\
&& + \; \pv{1}\ph{2}\ph{3}\pv{4} + \pv{1}\ph{2}\pv{3}\ph{4})/2, \nonumber \\
\eea
while the corresponding operator for three down-conversions 
occurring (one in PDC1, one in PDC2, and one in PDC3) is given by

\bea
\label{eq:threeDownConversions}
\hat{O}_{1,2}^\dagger\hat{O}_{3,4}^\dagger\hat{O}_{5,6}^\dagger 
&=& 
          (\ph{1}\pv{2} + \pv{1}\ph{2}) 
          (\ph{3}\pv{4} + \pv{3}\ph{4}) \nonumber \\
&&        (\ph{5}\pv{6} + \pv{5}\ph{6})/\sqrt{8} \nonumber \\
&=& 
          (\ph{1}\pv{2}\ph{3}\pv{4}\ph{5}\pv{6}   \nonumber \\
&&   
      + \;  \ph{1}\pv{2}\ph{3}\pv{4}\pv{5}\ph{6}   \nonumber \\
&&
      + \;  \ph{1}\pv{2}\pv{3}\ph{4}\ph{5}\pv{6}   \nonumber \\
&&
      + \;  \ph{1}\pv{2}\pv{3}\ph{4}\pv{5}\ph{6}   \nonumber \\
&&
      + \;  \pv{1}\ph{2}\ph{3}\pv{4}\ph{5}\pv{6}   \nonumber \\
&&
      + \;  \pv{1}\ph{2}\ph{3}\pv{4}\pv{5}\ph{6}    \nonumber \\
&&
      + \;  \pv{1}\ph{2}\pv{3}\ph{4}\ph{5}\pv{6}   \nonumber \\
&&
      + \;  \pv{1}\ph{2}\pv{3}\ph{4}\pv{5}\ph{6})\sqrt{8}.    \nonumber \\
\eea
If we now label the optical modes at the detector outputs of 
Fig.~\ref{fig:logical0} by  $\hat{a}^\dagger$, 
$\hat{b}^\dagger$, $\hat{c}^\dagger$, $\hat{d}^\dagger$, 
$\hat{e}^\dagger$, and $\hat{f}^\dagger$ and label 
the other three outputs by $\hat{g}^\dagger$, $\hat{h}^\dagger$, 
and $\hat{i}^\dagger$ the setup used to 
prepare the logical zero state can be seen 
in Fig.~\ref{fig:logical0modes} to transform the 
down-converted modes according to the following relations 
\bea
\label{eq:modeTransformations1}
\ph{1} &\rightarrow& \bv{1}, \;
\pv{1} \rightarrow \gv{1}, \nonumber \\
\ph{2} &\rightarrow& \hv{2}, \;
\pv{2} \rightarrow (\dhnew{2} + \ev{2})/\sqrt{2}, \nonumber \\
\ph{3} &\rightarrow& \gh{3}, \;
\pv{3} \rightarrow \av{3},  \nonumber \\
\ph{4} &\rightarrow& (\dhnew{4} - \ev{4})/\sqrt{2}, \;
\pv{4} \rightarrow \hh{4}, \nonumber \\
\ph{5} &\rightarrow& \ih{5}, \;
\pv{5} \rightarrow \cv{5}, \nonumber \\
\ph{6} &\rightarrow& \fh{6}, \;
\pv{6} \rightarrow \dv{6}.
\eea
Using these transformation relations the simultaneous down-conversion 
operator in Eq.~(\ref{eq:threeDownConversions}) can then be 
calculated to transform according to the equation below. 
If detectors are placed 
at the outputs of the optical modes $\hat{a}^\dagger$, 
$\hat{b}^\dagger$, $\hat{c}^\dagger$, $\hat{d}^\dagger$, 
$\hat{e}^\dagger$, and $\hat{f}^\dagger$ 
and each register a detection simultaneously the state post-selected 
in this way is a superposition of the states corresponding to the 
terms in Eq.~(\ref{eq:result1}) containing all six of these modes. 
The two terms satisfying this requirement are 
$-\bv{1}\dhnew{2}\av{3}\ev{4}\cv{5}\fh{6}$ and 
$\bv{1}\ev{2}\av{3}\dhnew{4}\cv{5}\fh{6}$ (both scaled by the same 
coefficient) which yield the logical zero state given by 
Eq.~(\ref{eq:polarizationEncodingZeroAndOne}). 

\begin{figure}[!ht]
\includegraphics[width=.50\textwidth]{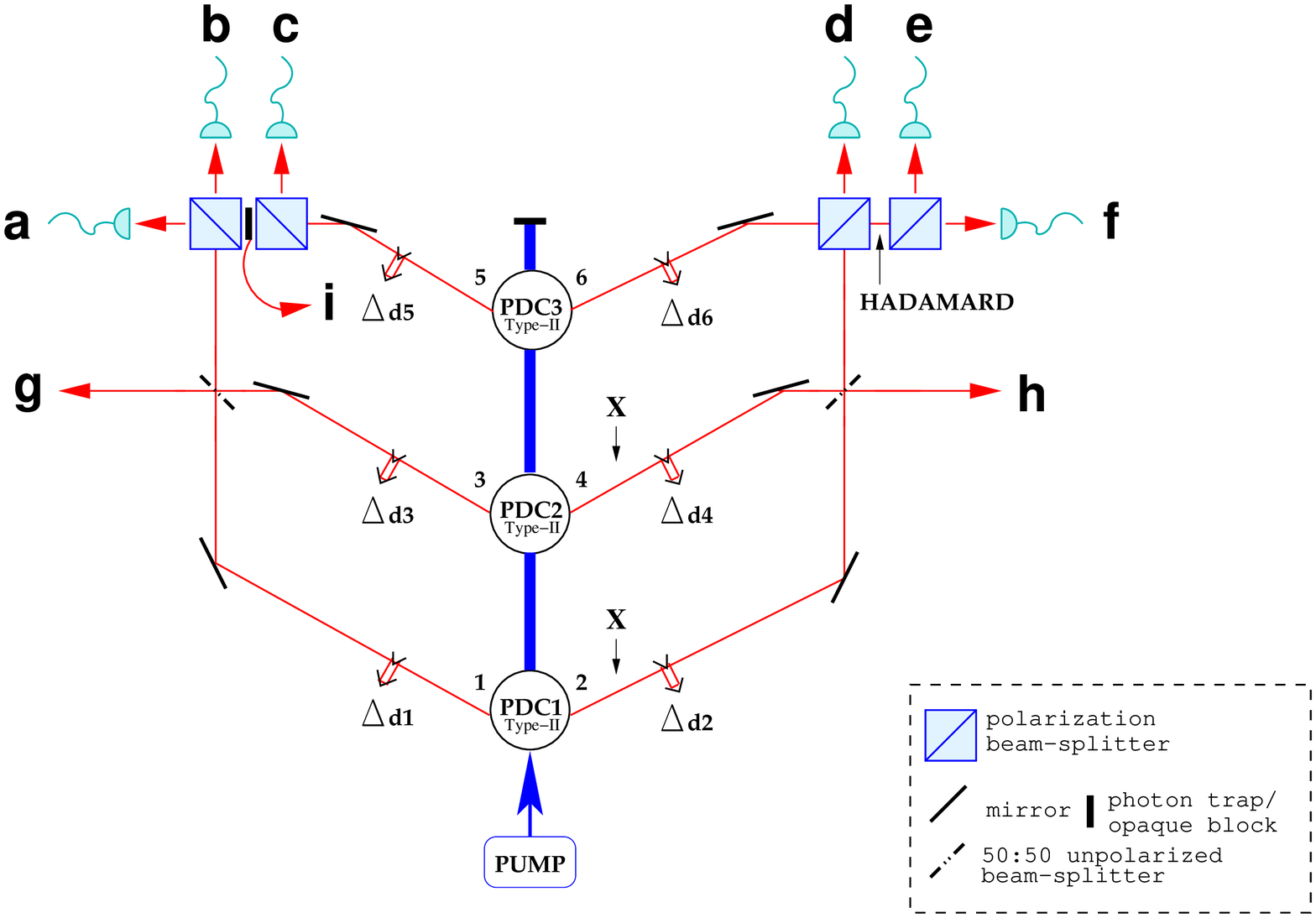}
\caption{(color online). An optical arrangement that may be used for the 
preparation of a decoherence-free logical one state.}
\label{fig:logical1modes}
\end{figure}

\bea
\label{eq:result1}
& &
\hat{O}_{1,2}^\dagger\hat{O}_{3,4}^\dagger\hat{O}_{5,6}^\dagger 
\rightarrow \nonumber \\ 
&&
 [\bv{1}(\dhnew{2} + \ev{2})(\frac{1}{\sqrt{2}})\gh{3}\hh{4}\ih{5}\dv{6} 
  \nonumber \\
&&
 + \; \bv{1}(\dhnew{2} + \ev{2})(\frac{1}{\sqrt{2}})\gh{3}\hh{4}\cv{5}\fh{6} 
  \nonumber \\
&&
 + \; \bv{1}(\dhnew{2} + \ev{2})(\frac{1}{\sqrt{2}})\av{3}(\dhnew{4} - \ev{4})
      (\frac{1}{\sqrt{2}})\ih{5}\dv{6} 
  \nonumber \\
&&
 + \; \bv{1}(\dhnew{2} + \ev{2})(\frac{1}{\sqrt{2}})\av{3}(\dhnew{4} - \ev{4})
      (\frac{1}{\sqrt{2}})\cv{5}\fh{6}
  \nonumber \\
&&
 + \; \gv{1}\hv{2}\gh{3}\hh{4}\ih{5}\dv{6} + \gv{1}\hv{2}\gh{3}\hh{4}\cv{5}
\fh{6}] \nonumber \\
&&
 + \; \gv{1}\hv{2}\av{3}(\dhnew{4} - \ev{4})(\frac{1}{\sqrt{2}})\ih{5}\dv{6}
  \nonumber \\
&&
 + \; \gv{1}\hv{2}\av{3}(\dhnew{4} - \ev{4})(\frac{1}{\sqrt{2}})\cv{5}\fh{6}
](\frac{1}{\sqrt{8}}) \nonumber \\ 
\eea
The setup shown in Fig.~\ref{fig:logical1modes}, used to 
prepare the logical one 
state, transforms the down-converted modes according to the 
following relations (ignoring the unimportant 
phase shifts in modes $\hat{g}^\dagger$ 
and $\hat{h}^\dagger$ ), Eq.~(\ref{eq:modeTransformations2})
\begin{widetext}
\bea
\label{eq:modeTransformations2}
\ph{1} &\rightarrow& (\gh{1} + \bh{1})\frac{1}{\sqrt{2}}, \;
\pv{1} \rightarrow   (\gv{1} + \av{1})\frac{1}{\sqrt{2}}, \;
\ph{2} \rightarrow (\sqrt{2}\hv{2} + \fh{2} - \ev{2})\frac{1}{2}, 
\nonumber \\
\pv{2} &\rightarrow& (\hh{2} + \dhnew{2})/\frac{1}{\sqrt{2}}, \;
\ph{3} \rightarrow (\gh{3} + \bh{3})\frac{1}{\sqrt{2}}, \;
\pv{3} \rightarrow (\gv{3} + \av{3})\frac{1}{\sqrt{2}},  \nonumber \\
\ph{4} &\rightarrow& (\sqrt{2}\hv{4} + \fh{4} - \ev{4})\frac{1}{2}, \;
\pv{4} \rightarrow (\hh{4} + \dhnew{4})\frac{1}{\sqrt{2}}, \;
\ph{5} \rightarrow \ih{5}, \nonumber \\
\pv{5} &\rightarrow& \cv{5}, \;
\ph{6} \rightarrow (\fh{6} + \ev{6})\frac{1}{\sqrt{2}}, \;
\pv{6} \rightarrow \dv{6}.
\eea
\end{widetext}
These relations are used below 
to calculate the transformation of Eq.~(\ref{eq:threeDownConversions}). 
Again, the terms containing each of the modes $\hat{a}^\dagger$, 
$\hat{b}^\dagger$, $\hat{c}^\dagger$, $\hat{d}^\dagger$, 
$\hat{e}^\dagger$, and $\hat{f}^\dagger$ are compatible with 
a sixfold coincidence detection. Expanding Eq.~(\ref{eq:result2}) reveals 
the four terms $\bh{1}\dhnew{2}\av{3}\fh{4}\cv{5}\ev{6}$, 
$-\bh{1}\dhnew{2}\av{3}\ev{4}\cv{5}\fh{6}$, 
$\av{1}\fh{2}\bh{3}\dhnew{4}\cv{5}\ev{6}$, and 
$-\av{1}\ev{2}\bh{3}\dhnew{4}\cv{5}\fh{6}$ (all scaled by the same 
coefficient) which meet this requirement. These terms correspond 
to the logical one state of Eq.~(\ref{eq:polarizationEncodingZeroAndOne}). 

\begin{widetext}
\bea
\label{eq:result2}
&&
\hat{O}_{1,2}^\dagger\hat{O}_{3,4}^\dagger\hat{O}_{5,6}^\dagger 
\rightarrow \nonumber \\
&& 
 [(\gh{1} + \bh{1})(\frac{1}{\sqrt{2}})(\hh{2} + \dhnew{2})(\frac{1}{\sqrt{2}})
(\gh{3} + \bh{3})(\frac{1}{\sqrt{2}}) 
+ 
(\hh{4} + \dhnew{4})(\frac{1}{\sqrt{2}})
\ih{5}\dv{6} \nonumber \\
&& 
+ \;
(\gh{1} + \bh{1})(\frac{1}{\sqrt{2}})(\hh{2} + \dhnew{2})(\frac{1}{\sqrt{2}})
(\gh{3} + \bh{3})(\frac{1}{\sqrt{2}}) 
+ 
(\hh{4} + \dhnew{4})(\frac{1}{\sqrt{2}})\cv{5}(\fh{6} + \ev{6})
(\frac{1}{\sqrt{2}}) \nonumber \\
&&
+ \;
(\gh{1} + \bh{1})(\frac{1}{\sqrt{2}})(\hh{2} + \dhnew{2})(\frac{1}{\sqrt{2}})
(\gv{3} + \av{3})(\frac{1}{\sqrt{2}}) 
+ 
(\sqrt{2}\hv{4} + \fh{4} - \ev{4})(\frac{1}{2})\ih{5}\dv{6} \nonumber \\
&&
+ \;
(\gh{1} + \bh{1})(\frac{1}{\sqrt{2}})(\hh{2} + \dhnew{2})(\frac{1}{\sqrt{2}})
(\gv{3} + \av{3})(\frac{1}{\sqrt{2}}) 
+ 
(\sqrt{2}\hv{4} + \fh{4} - \ev{4})(\frac{1}{2})\cv{5}(\fh{6} + \ev{6})
(\frac{1}{\sqrt{2}}) \nonumber \\
&&
+ \;
(\gv{1} + \av{1})(\frac{1}{\sqrt{2}})(\sqrt{2}\hv{2} + \fh{2} - \ev{2})
(\frac{1}{2}) 
+ 
(\gh{3} + \bh{3})(\frac{1}{\sqrt{2}})(\hh{4} + \dhnew{4})(\frac{1}{\sqrt{2}})
\ih{5}\dv{6} \nonumber \\
&&
+ \;
(\gv{1} + \av{1})(\frac{1}{\sqrt{2}})(\sqrt{2}\hv{2} + \fh{2} - \ev{2})
(\frac{1}{2}) 
+ 
(\gh{3} + \bh{3})(\frac{1}{\sqrt{2}})(\hh{4} + \dhnew{4})(\frac{1}{\sqrt{2}})
\cv{5}(\fh{6} + \ev{6})(\frac{1}{\sqrt{2}}) \nonumber \\
&&
+ \;
(\gv{1} + \av{1})(\frac{1}{\sqrt{2}})(\sqrt{2}\hv{2} + \fh{2} - \ev{2})
(\frac{1}{2}) 
+ 
(\gv{3} + \av{3})(\frac{1}{\sqrt{2}})(\sqrt{2}\hv{4} + \fh{4} - \ev{4})
(\frac{1}{2})\ih{5}\dv{6} \nonumber \\
&&
+ \;
(\gv{1} + \av{1})(\frac{1}{\sqrt{2}})(\sqrt{2}\hv{2} + \fh{2} - \ev{2})
(\frac{1}{2}) 
+ 
(\gv{3} + \av{3})(\frac{1}{\sqrt{2}})(\sqrt{2}\hv{4} + \fh{4} - \ev{4})
(\frac{1}{2})\cv{5}(\fh{6} + \ev{6})(\frac{1}{\sqrt{2}})](\frac{1}{\sqrt{8}}).
\nonumber \\
\eea
\end{widetext}
In Fig.~\ref{fig:singlet} we see that the mode transformations 
corresponding to the maximally entangled state arrangement are given by
\begin{widetext}
\bea
\label{eq:modeTransformations3}
\ph{1} &\rightarrow& \bh{1}, \;
\pv{1} \rightarrow \av{1}, \;
\ph{2} \rightarrow (\fv{2} + \dv{2} + \fh{2} + \ch{2})/2, \nonumber \\
\pv{2} &\rightarrow& (\fh{2} + \ch{2} + \sqrt{2}\ev{2})/2, \;
\ph{3} \rightarrow \ah{3}, \;
\pv{3} \rightarrow \bv{3},  \nonumber \\
\ph{4} &\rightarrow& (\hv{4} + \cv{4} + \hh{4} + \dhnew{4})/2, \;
\pv{4} \rightarrow (\hh{4} + \dhnew{4} + \sqrt{2}\gv{4})/2.
\eea
\end{widetext}
Using these transformation relations the simultaneous down-conversion 
operator in Eq.~(\ref{eq:twoDownConversions}) can then be 
calculated to transform according to Eq.~(\ref{eq:result3}) below.

If detectors are placed 
at the outputs of the optical modes $\hat{a}^\dagger$, 
$\hat{b}^\dagger$, $\hat{c}^\dagger$, and $\hat{d}^\dagger$ 
and each register a detection simultaneously the state post-selected 
in this way is a superposition of the states corresponding to the 
terms in Eq.~(\ref{eq:result3}) containing all four of these modes. 
The three terms satisfying this requirement are 
$\bh{1}\ch{2}\ah{3}\dhnew{4}$, $\av{1}\dv{2}\bv{3}\cv{4}$, and 
$\av{1}\ch{2}\bv{3}\dhnew{4}$ (all scaled by the same 
coefficient) which yield the state given by 
Eq.~(\ref{eq:singletPolarization}). 
\begin{widetext}
\bea
\label{eq:result3}
\hat{O}_{1,2}^\dagger\hat{O}_{3,4}^\dagger &\rightarrow& 
    [(\frac{1}{4})\bh{1}(\fh{2} + \ch{2} + \sqrt{2}\ev{2})\ah{3}(\hh{4} 
     + \dhnew{4} + \sqrt{2}\gv{4}) \nonumber \\
  &&   +  (\frac{1}{4})(\bh{1}(\fh{2} + \ch{2} + \sqrt{2}\ev{2})
        \bv{3}(\hv{4} + \cv{4} + \hh{4} + \dhnew{4})) \nonumber \\
  &&   + (\frac{1}{4})\av{1}(\fv{2} + \dv{2} + \fh{2} + \ch{2})
       \ah{3}(\hh{4} + \dhnew{4} + \sqrt{2}\gv{4})   \nonumber \\
  &&   + (\frac{1}{4})\av{1}(\fv{2} + \dv{2} + \fh{2} + \ch{2})
       \bv{3}(\hv{4} + \cv{4} + \hh{4} + \dhnew{4})] (\frac{1}{2}) 
\nonumber \\
\eea
\end{widetext}


\section*{Acknowledgments}

We greatfully acknowledge Hwang Lee for several helpful comments.  
This material is based upon work supported by the 
National Science Foundation under Grant No. 0545798 to MSB. 



\begin{thebibliography}{56}
\bibitem[{not()}]{note1}
\bibinfo{note}{Alternatively, we could use 
  $\ket{1} = 1/\sqrt{2}(\ket{VH}+\ket{HV})$ or $\ket{1} = \ket{HV}$.  In 
  any of these cases we are projecting onto a three-dimensional subspace 
  of a ququart (qudit with $d=4$).  The first case is one for which no 
  distinction is made between the first and second photons.  There is one 
  with each of the two polarizations.  Here we simplify our argument by 
  choosing one of the three projection possibilities, but this method can, 
  in principle, be modified to any of the three cases.}
\end{thebibliography}


\end{document}